\documentclass{aa}

\usepackage{graphicx}
\usepackage{txfonts}
\usepackage[dvipsnames]{xcolor}
\usepackage{xparse,xcolor}
\usepackage{ulem}
\begin{document}

   \title{Outbursts in ultracompact X-ray binaries}

   \author{J.-M. Hameury
          \inst{1}
          \and
          J.-P. Lasota\inst{2,3}
          }

   \institute{Observatoire Astronomique de Strasbourg, Université de Strasbourg, CNRS UMR 7550, 67000 Strasbourg, France\\
             \email{jean-marie.hameury@astro.unistra.fr}
         \and
         Institut d'Astrophysique de Paris, CNRS et Sorbonne Universit\'es, UPMC Paris~06, UMR 7095, 98bis Bd Arago, 75014 Paris, France
\and
             Nicolaus Copernicus Astronomical Center, Polish Academy of Sciences, Bartycka 18, 00-716 Warsaw, Poland           
}

   \date{}


  \abstract
   {Very faint X-ray binaries appear to be transient in many cases with peak luminosities much fainter than that of usual soft X-ray transients, but their nature still remains elusive.}
   {We investigate the possibility that this transient behaviour is due to the same thermal/viscous instability which is responsible for outbursts of bright soft X-ray transients, occurring in ultracompact binaries for adequately low mass-transfer rates. More generally, we investigate the observational consequences of this instability when it occurs in ultracompact binaries.}
   {We use our code for modelling the thermal-viscous instability of the accretion disc, assumed here to be hydrogen poor. We also take into account the effects of disc X-ray irradiation, and consider the impact of the mass-transfer rate on the outburst brightness.}
   {We find that one can reproduce the observed properties of both the very faint and the brighter short transients (peak luminosity, duration, recurrence times), provided that the viscosity parameter in quiescence is slightly smaller (typically a factor of between two and fours) than in bright soft X-ray transients and normal dwarf nova outbursts, the viscosity in outburst being unchanged. This possibly reflects the impact of chemical composition on non-ideal MHD effects affecting magnetically driven turbulence in poorly ionized discs.}
   {}


   \maketitle
%

\section{Introduction}

Soft X-ray transients (SXTs) are a subclass of low mass X-ray binaries (LMXBs) which alternate between quiescent periods lasting years and bright outbursts which can reach the Eddington limit and last typically for weeks \citep[see e.g.][for reviews of the observations]{ts96,csl97,yy15}. These bright sources were discovered soon after the first X-ray satellites came into operation. With the advent of more sensitive X-ray telescopes such as \textit{Chandra}, \textit{XMM-Newton} and \textit{Swift}, fainter and fainter sources were discovered \citep[see e.g.][]{hww04,mpb05,pgb05,swd05} during observations of galactic fields and particularly the galactic centre. \citet{wzr06} then introduced two new classes of transient sources, the so-called faint X-ray transients, and very faint X-ray transients (VFXTs) that have peak luminosities in the range 10$^{36-37}$ erg s$^{-1}$, and 10$^{34-36}$ erg s$^{-1}$ respectively.  These VFXTs are part of the very faint X-ray binaries group (VFXBs) which are mostly transients with quiescent luminosities below 10$^{33}$ erg s$^{-1}$ \citep{hbd15}. Despite the fact that they were discovered more than ten years ago, their nature still remains elusive, probably in part because of their faintness, and also because most of them are located in highly absorbed regions making the identification of optical counterparts difficult. It is in particular not clear that the groups of faint and very faint transients are homogeneous, and searching for a model accounting for properties of all VFXBs might be inappropriate.

It is widely accepted that the transient nature of the classical soft X-ray transients (SXTs), which are  much brighter in outburst than VFXBs and can reach luminosities of order of the Eddington limit, is due to the thermal/viscous instability of the accretion disc in regions where its effective temperature is of order of 6,000 -- 8,000 K \citep{kr98,dhl01}. This occurs when the mass-transfer rate from the secondary lies in some range $[\dot{M}_{\rm crit}^-(r_{\rm in}) - \dot{M}_{\rm crit}^+(r_{\rm out})]$ \citep{lasota-01}, where $r_{\rm in}$ and $r_{\rm out}$ are the inner and outer disc radius respectively, and $\dot{M}_{\rm crit}^-$ and $\dot{M}_{\rm crit}^+$ are the critical mass-transfer rates for the instability to occur at a given point in the disc ($\dot{M}_{\rm crit}^-$ is the maximum value for the disc to stay on the cold, stable branch, and $\dot{M}_{\rm crit}^+$ is the minimum value for the disc to stay on the hot branch). The observed luminosities of persistent and transient X-ray binaries show that this in indeed the case \citep{vp96,cfd12} if the accretion disc is X-ray self-irradiated. 

It is tempting to assume that faint and very faint outbursts could be due to the same instability that is responsible for the bright outbursts of SXTs. Such a model has to explain the faintness of the outbursts and take into account the low average mass accretion rate, which for VFXBs lies in the range $3 \times 10^{-13}$ -- $1 \times 10^{-10}$ M$_\odot$ yr$^{-1}$ \citep{dw09}. This low mass-transfer rate is directly related to the nature of the secondary and its evolutionary status, and, to a lesser extent, to the primary mass. 
\citet{k00} suggested that faint transients contain low mass secondaries that have evolved past the minimum orbital period. \citet{kw06} proposed later that VFXBs were either formed with a brown dwarf or a planetary companion, or could be the end products of massive primordial binaries which lead to systems containing an intermediate mass black hole, of the order of 1000 M$_\odot$. The latter explanation would of course not hold for VFXBs that exhibit type I X-ray bursts, and must therefore contain a neutron star. As noted by \citet{hbd15}, many of VFXBs have exhibited type I X-ray bursts, for example XMM J174457-2850.3 \citep{dwr14} and therefore do not harbor a black hole. \citet{mp13} proposed that the weak mass-transfer rates in VFXBs can be explained if these systems are detached, being in a similar state as cataclysmic variables in the period gap. This occurs when the secondary star becomes fully convective as a result of mass loss; angular momentum losses are reduced and the slightly bloated secondary shrinks, returning to thermal equilibrium; mass transfer can only occur via the faint secondary wind. The mass-transfer rate is difficult to estimate, but \citet{mp13} quote values of order of $10^{-16} - 10^{-14}$ M$_\odot$ yr$^{-1}$, which are much lower than the values of the order of $10^{-13}$ M$_\odot$yr$^{-1}$ 
inferred from observations of VFXTs \citep{dw10}.

If the transient nature of the faint X-ray sources and VFXBs were to be attributed to the thermal/viscous disc instability, the observational difference between bright transients and VFXBs would have to be due either to a small accretion disc for the latter, (the critical mass-transfer rates $\dot{M}_{\rm crit}^+$ and $\dot{M}_{\rm crit^-}$ and the outburst peak luminosity scale as $r^{2.65}$), and/or to the fact that only a small fraction of the disc is involved in the instability. The smallest accretion discs are found in ultracompact binaries. These systems form a subclass of low-mass X-ray binaries (LMXBs) with very short orbital periods (typically less than $\sim$ 1hr). Unless the system has followed a very special evolutionary track, the secondary star is hydrogen poor in order to fit inside its Roche lobe, being either a degenerate helium star or the core of a carbon-oxygen or oxygen-neon-magnesium white dwarf \citep[see e.g.][]{N08}. Compared to hydrogen-rich binaries, this change in chemical composition significantly increases the temperature for which the disc becomes unstable, and hence the critical mass-transfer rates, but the global picture remains unchanged \citep{ldk08}.

In this paper, we investigate in detail the nature and observational consequences of the disc instability model (DIM) applied to ultracompact binaries, in particular taking  into account the effects of self-irradiation by accretion-produced X-rays. \cite{ldk08} estimated the stability conditions for hydrogen deficient ultracompact binaries, taking into account these effects, but have not produced light curves that can be directly compared with observations. This is needed in particular when one is interested in the faint end of the outburst luminosity distribution, for which the outermost parts of the disc may remain in the cold state during an outburst, and the peak luminosity is far from reaching the maximum luminosity one can derive using simple analytical estimates. \cite{kld12} did calculate the light curves predicted by the DIM in the case of degenerate helium secondaries, but, as they were interested in AM CVn stars, they did not consider disc irradiation. We show here that the observational properties (peak luminosities, duration, recurrence time) of many of the VFXB transients can be accounted for by the DIM provided, however, that the viscosity in quiescence is slightly smaller (typically a factor of between two and four) than in hydrogen--rich binaries. This does not imply that all VFXBs are ultracompact binaries -- some, such as for example \object{AX J1745.6-2901} which has an orbital period of 8.4 hr, are obviously not. Other systems, such as \object{1RXH J173523.7-354013} for which an H$\alpha$ line has been detected \citep[][note that this particular object is a persistent source, with an X-ray luminosity of order of $2 \times 10^{35}$ erg s$^{-1}$]{djt10}  may contain a hydrogen rich secondary and are therefore not ultracompact. It does, however, mean that the short and faint outbursts observed in a number of VFXTs can be explained by the DIM when applied to ultracompact binaries. 

We leave for future work the much needed application of the DIM to the case where the donor star is a carbon-oxygen degenerate star, and in particular the confirmation that the C/O discs would behave in a manner similar to hydrogen rich discs.

\section{DIM for irradiated helium accretion discs}

\subsection{Vertical structure}

The effect of X-ray irradiation on the local disc vertical structure at a given radius $r$ is merely to change the surface boundary condition:
\begin{equation}
T_{\rm surf}^4 = T_{\rm eff}^4 + T_{\rm irr}^4,
\end{equation}
where $T_{\rm eff}$ is the effective temperature corresponding to the heat flux generated by viscous dissipation is the disc, $T_{\rm surf}$ is the temperature at the disc surface, defined as the place where the optical depth is 2/3, and  $T_{\rm irr}$ is the local irradiation temperature at point $r$ given by
\begin{equation}
\sigma T_{\rm irr}^4 = C \frac{L_{\rm X}}{4 \pi r^2},
\end{equation}
with
\begin{equation}
L_{\rm X} = \epsilon \dot{M}_{\rm in} c^2.
\end{equation}
$\dot{M}_{\rm in}$ is the mass accretion rate at the inner disc radius, $\epsilon$ is the accretion efficiency, typically of order of 0.1. The coefficient $C$ measures the fraction of the X-ray luminosity which is intercepted by the disc; it could be in principle calculated in a self consistent way, but this turns out to be a formidable task. The coefficient $C$ also incorporates the fact that only a fraction of the X-rays is absorbed below the photosphere and is hence able to affect the thermal structure of the disc; X-rays absorbed in the upper photosphere would rather be responsible for the formation of a hot corona whose interaction with the underlying cooler layers is complex. Based on observations, \cite{dhl01} showed that with $C = 5 \times 10^{-3}$, the model reproduces reasonably well the light curves of SXTs. It has been shown, however, that models are rather insensitive to the value of C, as long as $5\times 10^{-3} \lesssim C \lesssim 2\times 10^{-2}$ \citep[see][but we note that in this paper $C$ contains accretion efficiency $\epsilon$]{ldk08}, so for convenience we will assume in this paper that $C=0.01$.

\begin{figure}
   \centering
   \includegraphics[angle=-90,width=\columnwidth]{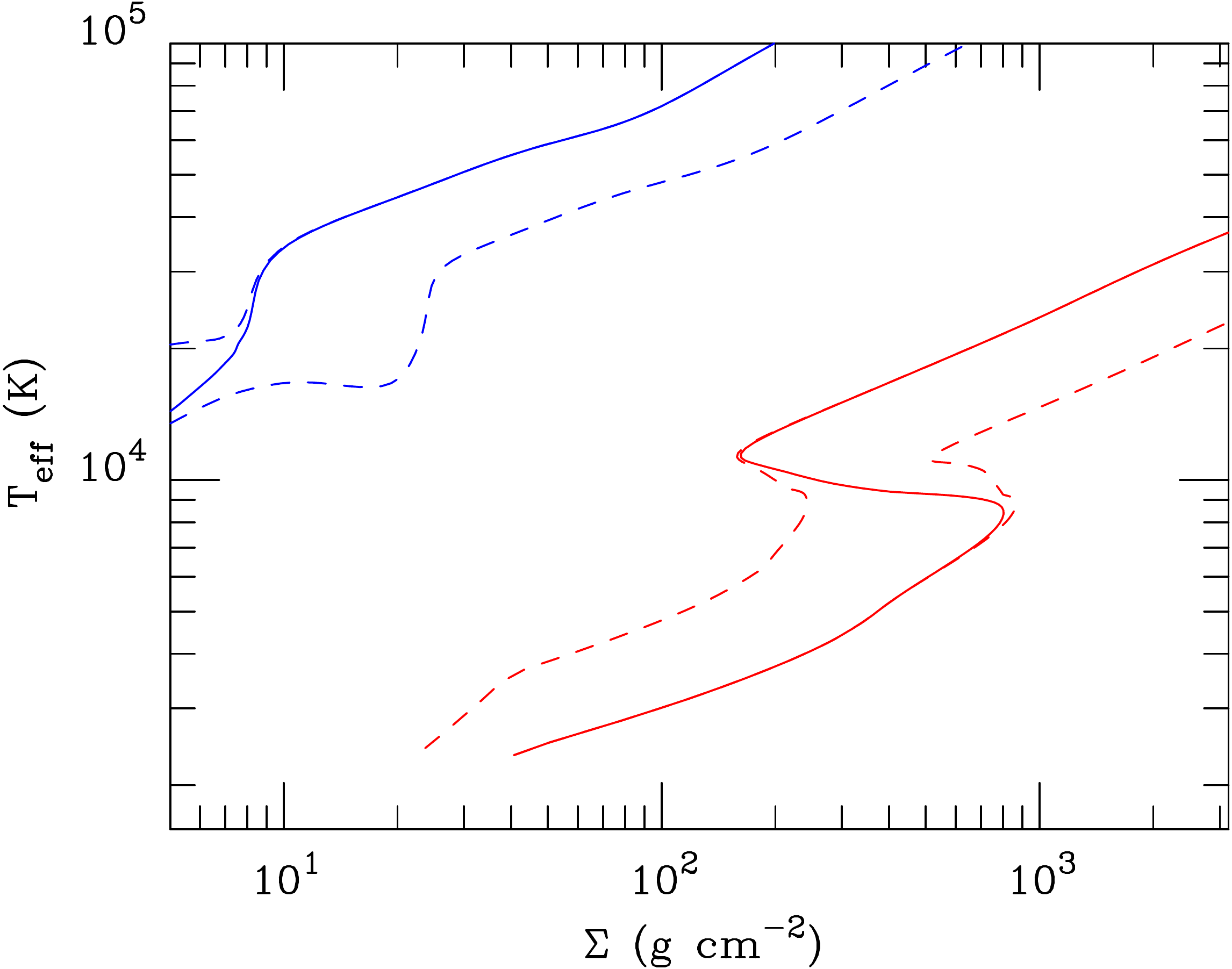}
   \caption{Two examples of local S curves -- effective temperature vs. surface density -- in a He disc when irradiation is included. The upper blue set of curves shows the case $T_{\rm irr} = 25,000$ K and $r=5 \times 10^{8}$ cm, the lower red set of curves corresponds to $T_{\rm irr} = 5,000$ K and $r= 10^{10}$ cm. In both cases, the solid curve corresponds to $\alpha$ given by Eq. (\ref{alpha}), while the dotted curves are for $\alpha = 0.05$ and $\alpha = 0.2$ (rightmost and leftmost curves respectively). The primary mass is 1.4 M$_\odot$. }
   \label{scurve}%
   \end{figure}

The disc vertical structure equations have been solved on a grid of parameters : surface density $\Sigma$, central disc temperature $T_{\rm c}$, radius $r$ and irradiation temperatures $T_{\rm irr}$, as described in \cite{ldk08}. Figure \ref{scurve} (dashed lines) shows the classical helium-disc thermal equilibrium S-curves. They have the same behaviour as in the hydrogen case, except that the effective temperatures of the turning points are higher, because the ionization potential of helium is larger than that of hydrogen.

As for cataclysmic variables and low-mass X-ray binaries, we assume that the Shakura-Sunyaev parameter $\alpha$ is a function of the disc central temperature, so that it is constant and equal to $\alpha_{\rm h}$ on the hot stable branch, and to $\alpha_{\rm c}$ on the cool lower branch; to smoothly connect the two branches, we take:
\begin{equation}
\log(\alpha) = \log(\alpha_{\rm c}) + \left[ \log(\alpha_{\rm h}) - \log(\alpha_{\rm c})\right] \times \left[ 1 + \left(\frac{4 \; 10^4 \; \rm K}{T_{\rm c}} \right)^{10}\right]^{-1}
\label{alpha}
\end{equation}
where $T_{\rm c}$ is the disc mid-plane temperature. The choice of this function is rather arbitrary, and is found to have little influence on our results, provided that the two stable branches including the turning points are unaffected; it only impacts the detailed structure of the transition fronts. This relation is similar to the one used in \cite{hmdlh98}, except that we now have a transition temperature of 40,000K instead of 25,000 K, reflecting the higher temperatures found for the turning points. Note also that the exponent for the temperature dependence is slightly different from the one in \citet{hmdlh98} (10 instead of 8). This change was dictated by the necessity to have a sharp enough transition between the hot and cold branches, to avoid affecting the stable parts of these branches, and an index of 8 would have resulted in a cold-branch maximum $\Sigma$ value on the the S curve obtained using Eq. (\ref{alpha}) slightly smaller than the value obtained with a constant $\alpha$ equal to $\alpha_{\rm c}$. Figure \ref{scurve} shows that with our choice for the transition temperature, the resulting S curve does indeed correspond to the case $\alpha = \alpha_{\rm c}$ on the cool branch, and $\alpha = \alpha_{\rm h}$ on the hot branch. It can also be seen that such a choice does not artificially generates unstable branches when $T_{\rm irr}$ is large enough to stabilize the disc. Finally note that whereas the effective temperature of both turning points is weakly dependent on radius, the disc mid-plane temperature is practically constant \citep[see the analytical fits in][]{ldk08}, explaining why the temperature scaling in Eq. (\ref{alpha}) can be taken to be independent of $r$. In the DIM, an \textit{ad hoc} jump in $\alpha$'s value is assumed in order to reproduce disc outbursts properties, their amplitude in particular. It has been found recently by \citet{coleman16} that intermittent convection appearing near the ionization temperature increases the viscosity parameter by the amount required by the DIM, thus providing a physical mechanism for the ansatz used. 

\subsection{Time evolution}

The maximum outburst luminosity is obtained assuming that during the outburst, the disc is close to a steady state with a mass accretion rate equal to the critical value $\dot{M}_{\rm crit}^+$ for the disc to stay on the hot stable branch \citep{ldk08}:
\begin{equation}
L_{\rm max} \simeq 3.5 \times 10^{37} \left( \frac{P_{\rm orb}}{1 \; \rm h} \right)^{1.67} \; \rm erg \; s^{-1},
\label{lmax}
\end{equation}
in the case of pure helium discs, assuming a 1.4 M$_\odot$ primary; $L_{\rm max}$ is almost independent of $\alpha_{\rm h}$. The upper limit is reached when the heating front is able to propagate up to the outer edge of the disc; for small values of the mass-transfer rate, this need not be the case, as the outer part of the disc could remain  on the cold, stable branch even during outbursts. The outer parts of unirradiated discs may remain cold when the mass-transfer rate is less than $\dot{M}_{\rm crit}^-$. However, irradiation enables the heating front to propagate further away during outbursts; using Eq. (\ref{lmax}), the peak irradiation temperature at the outburst maximum can reach:
\begin{equation}
T_{\rm irr, max} = 37,000 \; \left( \frac{P_{\rm orb}}{1 \; \rm h} \right)^{1/3} \left( \frac{C}{10^{-2}} \right)^{1/4} \; K,
\end{equation}
where we have assumed that the disc radius is 0.6 times the orbital separation \citep{p77}, and the primary mass is that given above. The maximum irradiation temperature at the outer edge of the disc is comparable, albeit slightly smaller than the critical temperature of 40,000 K at which the transition between $\alpha_{\rm c}$ and $\alpha_{\rm h}$ occurs. This significantly softens the condition $\dot{M}_{\rm tr} > \dot{M}_{\rm crit}^-$ as a condition for the heating front to reach the outer disc edge.

One therefore expects that systems with orbital periods close to one hour and relatively large mass-transfer rates will undergo bright outbursts approaching the Eddington limit, and thus not very different from those found in standard SXTs; systems with shorter orbital periods and/or low mass-transfer rates should exhibit only faint outbursts.

\begin{figure}
   \centering
   \includegraphics[width=\columnwidth]{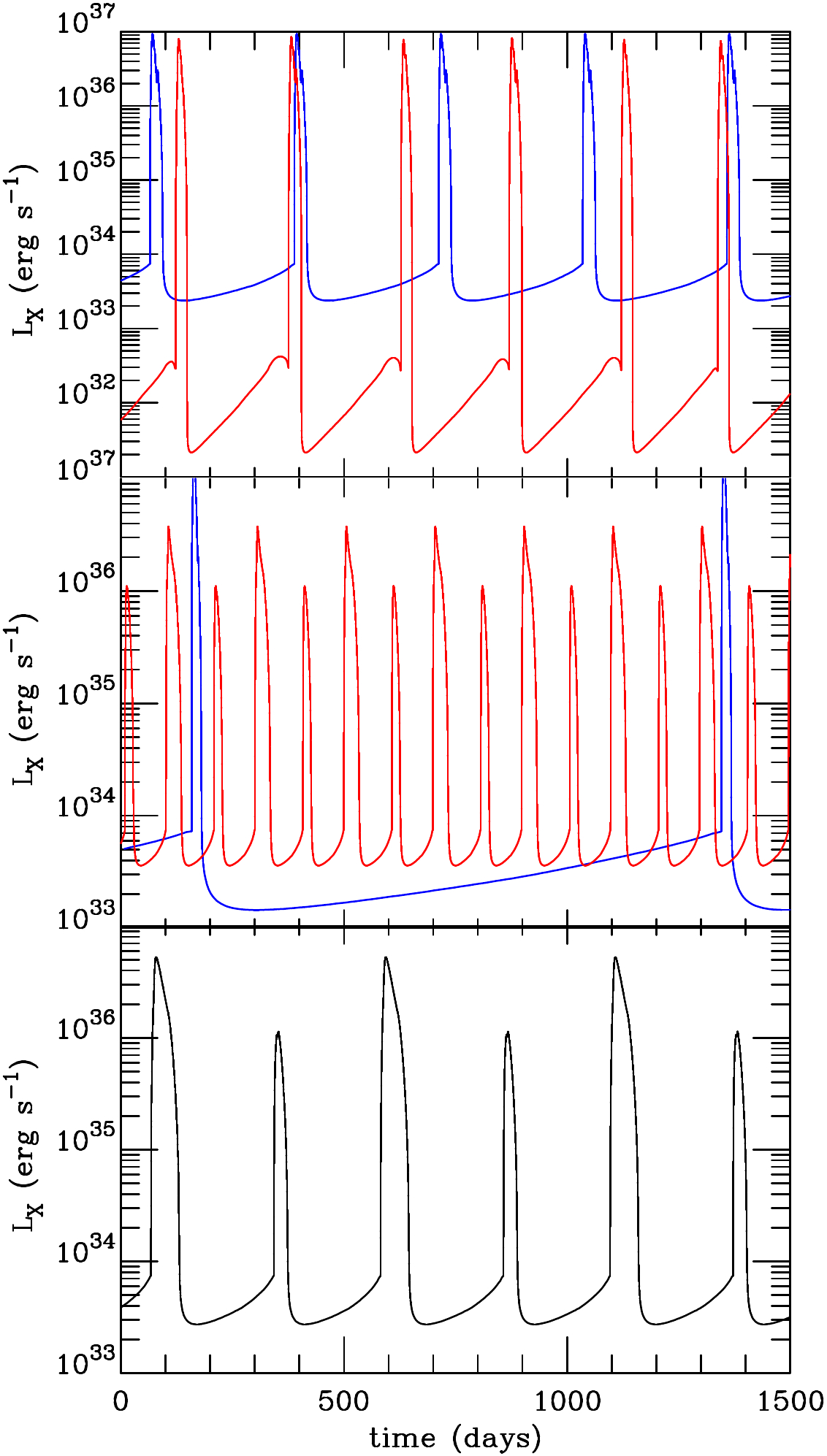}
   \caption{X-ray luminosity of an ultracompact transient system with a He disc, for different values of the viscosity parameters $\alpha_{\rm c}$ and $\alpha_{\rm h}$. Top panel: $\alpha_{\rm c} = 0.02$ and $\alpha_{\rm h}=0.2$; the  blue curve shows the case where the inner disc is truncated at a variable radius with time, according to Eq. (\ref{eq:rtrunc}), and the red curve is for a fixed inner radius ($2 \times 10^8$ cm). The mass-transfer rate is $3 \times 10^{15}$ g s$^{-1}$. Intermediate panel: changing $\alpha_{\rm c}$. Same as above for a truncated disc, with $\alpha_{\rm c} = 0.04$ (red curve) and $\alpha_{\rm c}=0.01$ (blue curve); $\alpha_{\rm h}=0.2$ is unchanged. In the case of the blue curve, the mass-transfer rate has been reduced by a factor 3. Bottom panel: changing $\alpha_{\rm h}$. Same as top panel, with $\alpha_{\rm h}=0.1$. $\alpha_{\rm c} = 0.02$ is unchanged}
   \label{lc-He}%
   \end{figure}
   
   \begin{figure}
   \centering
   \includegraphics[width=\columnwidth]{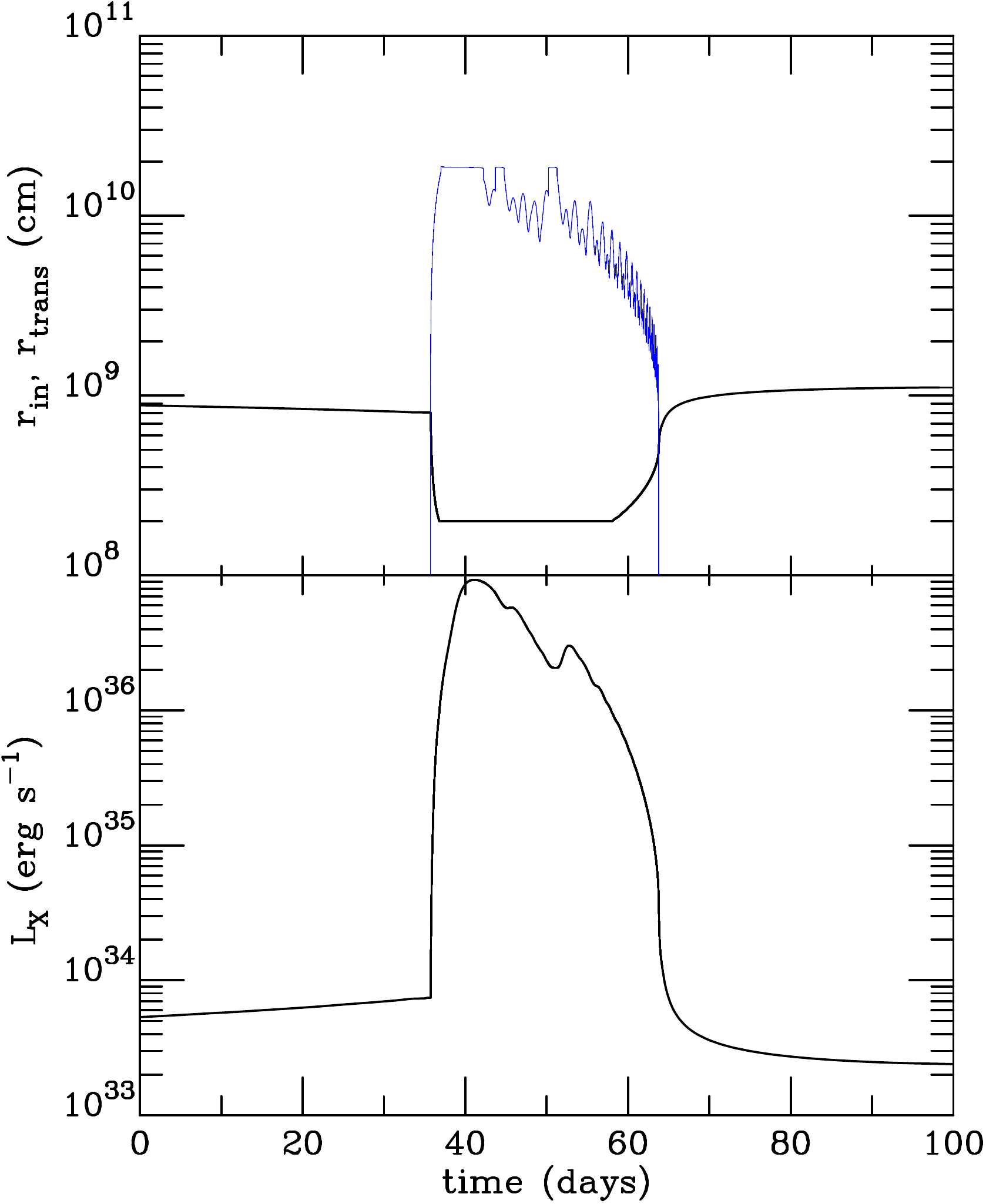}
   \caption{Detailed outburst profile for system shown in the upper panel of Fig. \ref{lc-He}, with truncation of the inner disc at a variable radius. The X-ray luminosity is shown in the lower panel, and the upper panel shows the inner disc radius (solid black curve) and the position of the transition front (thin solid blue line).}
   \label{lc-He-outburst}%
   \end{figure}
   
   In order to follow the time evolution of these systems, we use the numerical scheme of \cite{hmdlh98}, modified to include heating by irradiation as described in \citet{dhl01}. We have assumed that either the inner disc radius is set to a fixed, small value (ideally, this should be the radius of the neutron star or that of the innermost stable orbit for a black hole or a very compact neutron star; this is numerically not possible, and we instead typically take $r_{\rm in} = 2 \times 10^8$ cm), or assume, as in the case of standard bright X-ray transients, that it is truncated at a value which depends on the accretion rate at the inner edge. In this case, we set, unless otherwise noted:
  \begin{equation}
  r_{\rm in} = 2.0 \times 10^8  \left( \frac{\dot{M}_{\rm acc}}{10^{16} \; \rm g s^{-1}} \right)^{-2/7} \; \rm cm,
  \label{eq:rtrunc}
  \end{equation}
which is formally identical to the case of truncation by a magnetic dipolar field, with a magnetic dipole moment of $2 \times 10^{29}$ Gcm$^3$; one should note however that the magnetic moment of a neutron star is not large enough to disrupt the accretion flow at those large distances; truncation would have to be caused by another mechanism such as evaporation of the inner disc. Whatever the reason, the detailed functional dependence of $r_{\rm in}$ versus $\dot{M}_{\rm acc}$ is of little importance \citep{mhln00}: what matters is the value of $r_{\rm in}$ during quiescence. 

The outer disc radius is determined by the tidal torques, which can be written as
\begin{equation}
T_{\rm tid} = c_{\rm tid} \Omega_{\rm orb} \nu \Sigma f(r/a),
\end{equation}
where $c_{\rm tid}$ is a constant, $\Omega_{\rm orb}$ the orbital frequency, $\nu$ the viscosity, $\Sigma$ the surface column density, $a$ the orbital separation and $f$ a function describing the radial dependence of the torque. Several prescriptions can be used for $f$; based on \citet{pp77}, \citet{hmdlh98} used $f(r/a)=(r/a)^5$. This prescription has the disadvantage of allowing the outer disc radius to increase to large values, possibly beyond the Roche lobe. 
One could also assume that the tidal torques are negligibly small as long as the disc stays within the tidal truncation radius $r_{\rm tid}$, so that $T_{\rm tid} (r) =0$ for $r < r_{tid}$, and become arbitrarily large for $r>r_{\rm tid}$, thereby ensuring that the radius does not exceed $r_{\rm tid}$. This approach was chosen by \citet{vh08}. For practical numerical reasons, we have chosen instead:
\begin{equation}
f(r/a)=e^{K(r-r_{\rm tid})/a}
\end{equation}
where $K$ is a constant. This formulation is identical to \citet{vh08} when $K$ is large. Here, we take $K=50$ which is numerically convenient and ensures that the outer disc radius does not exceed $r_{\rm tid}$ by more than  a couple of percent. We did check that our numerical results do not depend significantly on the assumed value for $K$, provided that $K$ is not small. In this way, the accretion disc does not extend significantly beyond the tidal  truncation radius, but is allowed to be smaller. This prescription is therefore very different from assuming that the outer disc radius is fixed, which is unphysical and leads to light curves which are quite different from those with variable outer disc radii \citep{hmdlh98}.

\subsubsection{Bright outbursts}   
   Figure \ref{lc-He} (top panel) shows the time evolution of a system with orbital period 0.7 hr, primary mass 1.4 M$_\odot$, tidal truncation radius 1.8 $\times$ 10$^{10}$ cm, with and without truncation of the inner disc. The viscosity parameter $\alpha$ is taken to be 0.02 in the cold state, and 0.2 in the hot state. The mass-transfer rate is 3 $\times$ 10$^{15}$ g s$^{-1}$, and we have used $C=0.01$. As can be seen, truncation does not significantly affect the outburst properties of the system, in stringent contrast with standard SXTs with longer orbital periods. The recurrence time is slightly reduced when $r_{\rm in}$ is kept fixed to a small value, by about 25\%, the peak luminosity is also reduced (by about 20\%), and the outburst duration is unchanged. The most noticeable difference is the quiescence luminosity, which is approximately constant at $L_{\rm X} \sim 10^{33}$ erg s$^{-1}$ in the case of disc truncation and may be as low as $10^{31}$ erg s$^{-1}$ otherwise, with significant variations between outbursts. This difference is simply due to the fact that in quiescence, the mass accretion rate onto the compact object must be smaller than $\dot{M}_{\rm crit}^-(r_{\rm in})$, which rapidly increases with the truncation radius.
   
These orbital parameters as well as the mass-transfer rate are appropriate for the case of for example \object{XTE J1751-305}, which is a transient accreting millisecond pulsar whose 42 minutes orbital period has been determined from the Doppler variation of the spin period of the neutron star \citep{mss02}. We find that the outburst properties are not very different from the observed ones \citep[peak luminosity of order of $2 \times 10^{37}$ erg s$^{-1}$ for a distance of 2 kpc, outburst duration $\sim$ 10--15 days;][]{mss02}. This source underwent four outbursts over a timespan of 12 years \citep{hjw09}; the recurrence time produced by our model is of order of 1 yr, too short by a factor of three. Longer recurrence times can be obtained with smaller values of $\alpha_{\rm c}$; we found that with $\alpha_{\rm c} = 0.01$, one obtains a recurrence time of order of 3 years; the outburst peak luminosity is also increased by a factor of three, its duration being almost unaffected. Reducing the mass-transfer rate by a factor od three results in the light curve shown in the intermediate panel of Fig.\ref{lc-He} (blue curve). One should note that the ratio $\alpha_{\rm h}/\alpha_{\rm c}$ is at least of the order of ten, and larger by a factor of between roughly two and four. than the one used in models of dwarf novae or longer period transient LMXBs. One need to have a relatively low value of $\alpha_{\rm c}$ in order to account for the observed recurrence times, of the order of one year or more in ultracompact binaries, whereas $\alpha_{\rm h}$ cannot be too small in order to obtain outbursts whose duration does not exceed by far the observed values. 
   
 The intermediate and lower panel of Fig. \ref{lc-He} illustrate the effects of changing the viscosity. As mentioned above, the main impact of changing in $\alpha_{\rm c}$ is on the recurrence time: increasing (decreasing) $\alpha_{\rm c}$ by a factor of two results in a decrease (increase) of the recurrence time by a factor of three, larger than the relative change in $\alpha_{\rm c}$ because the viscosity is proportional to $\alpha c_{\rm s} H$, $c_{\rm s}$ and $H$ being the mid-plane sound speed and the vertical scale height respectively, which both decrease when $\alpha$ (and hence the central temperature) is reduced. Figure \ref{lc-He} also shows that modifying $\alpha_{\rm h}$ changes the duration of the outburst \citep[see e.g.,][]{KL12}, with relatively little influence on the recurrence time. The duration of the large outbursts shown in the bottom panel of Fig. \ref{lc-He} is of order of 60 days, much longer than the observed duration of XTE J1751-305 outbursts. It should finally be noted that when the ratio $\alpha_{\rm h}/\alpha_{\rm c}$ becomes too large, a sequence of alternating long and short outbursts appear. In the case  $\alpha_{\rm c} = 0.04$,  $\alpha_{\rm h} = 0.2$ corresponding to the case of classical dwarf novae, a sequence of bright and faint outbursts is observed, with a short recurrence time (100 days, 200 days between bright outbursts), as expected. 
      
   Figure \ref{lc-He-outburst} shows the outburst profile in the case of disc truncation, as well as the position of the transition front. It should be stressed out that the heating and cooling fronts are not as steep as in the standard (i.e. low mass primary as in CVs, and weak or no disc irradiation) model; such broad transition fronts are also found in the standard SXT disc instability model \citep[see e.g. Fig. 8 in][]{dhl01}. When the first cooling front forms, it is  quite narrow, but cannot propagate down to the inner disc edge, and is reflected into a heating front that propagates outwards. Several such reflections occur during which the width of the cooling front gradually increases. This happens because irradiation has a strong impact on the disc temperature $T_{\rm c}$, since $T_{\rm c}$ has to be larger than $T_{\rm irr}$ and can be very close to this value. But $T_{\rm irr}$ is a non local quantity, given by the accretion rate at the inner disc edge, and the radial temperature profile is largely determined by $T_{\rm irr}$. This is generic to strongly irradiated disc. On the other hand, for values of $C$ smaller than the ones considered in this paper, typically $C \leq 5 \times 10^{-4}$, this would not happen and the cooling front would be narrow. Here we define the transition radius as the point where $\alpha = (\alpha_{\rm c} + \alpha_{\rm h}) /2$.
   
One can note oscillations in the position of the transition front, as is also the case in long period SXTs \citep{dhl01}. These oscillations are here barely noticeable in the light curve so practically unobservable as such, but see below. The heating front propagates throughout the entire disc, which remains fully in the hot state for a few days before a cooling front can start propagating from the outer edge. This cooling front is unable to propagate down to the innermost radius, but is instead reflected into a heating front; the succession of heating and cooling fronts makes the decay much longer than if a cooling front were propagating freely. Interestingly, the decay is very roughly exponential, even during the phases where the disc is not kept in a hot state by irradiation. This shows that outbursts with exponential decays need not be produced according to \cite{kr98} model, as is sometimes incorrectly implied in the literature: \citet{kr98} showed that exponential decays are obtained during phases where the whole disc remains illuminated, on the hot branch and the cooling front propagation is quenched, while linear decays occur when the cooling front propagates freely. We have shown, however, that in many cases, when the cooling front propagation is controlled by irradiation, the decay from outburst is still close to exponential. Finally, it is worth noting that a significant fraction (about 30\%) of the disc mass is accreted during an outburst.
  
\subsubsection{Very faint outbursts}

\begin{figure}
   \centering
   \includegraphics[angle=-90,width=\columnwidth]{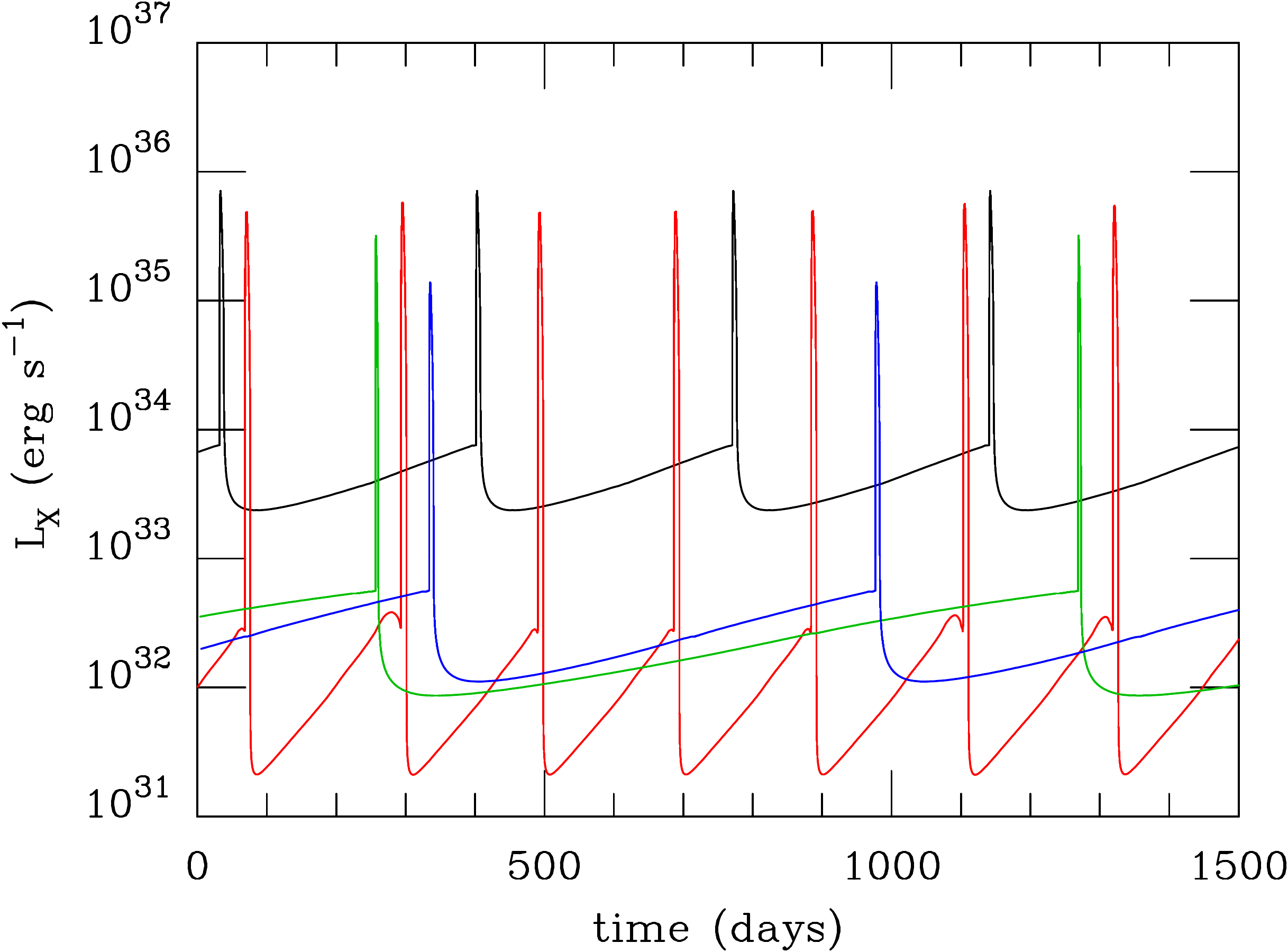}
   \caption{Light curves for low mass-transfer rates. For the black and red curves, $\dot{M}_{\rm tr} = 10^{14}$ gs$^{-1}$, the orbital parameters and the viscosity being identical to those of Fig. \ref{lc-He}. The black line shows the case of a truncated disc with an inner disc radius given by Eq. (\ref{eq:rtrunc}). The red curve corresponds to a fixed inner radius of 2 $\times$ 10$^8$ cm. The green and blue curves correspond to  $\dot{M}_{\rm tr} = 10^{13}$ gs$^{-1}$, the inner radius being equal to 0.15 times the value given by Eq. \ref{eq:rtrunc}. The viscosity parameters are $\alpha_{\rm h}=0.2$. $\alpha_{\rm c} = 0.01$ (green curve) and $\alpha_{\rm h}=0.1$. $\alpha_{\rm c} = 0.01$ (blue curve).}
   \label{lc_vfxb}%
   \end{figure}

\begin{figure}
   \centering
   \includegraphics[angle=-90,width=\columnwidth]{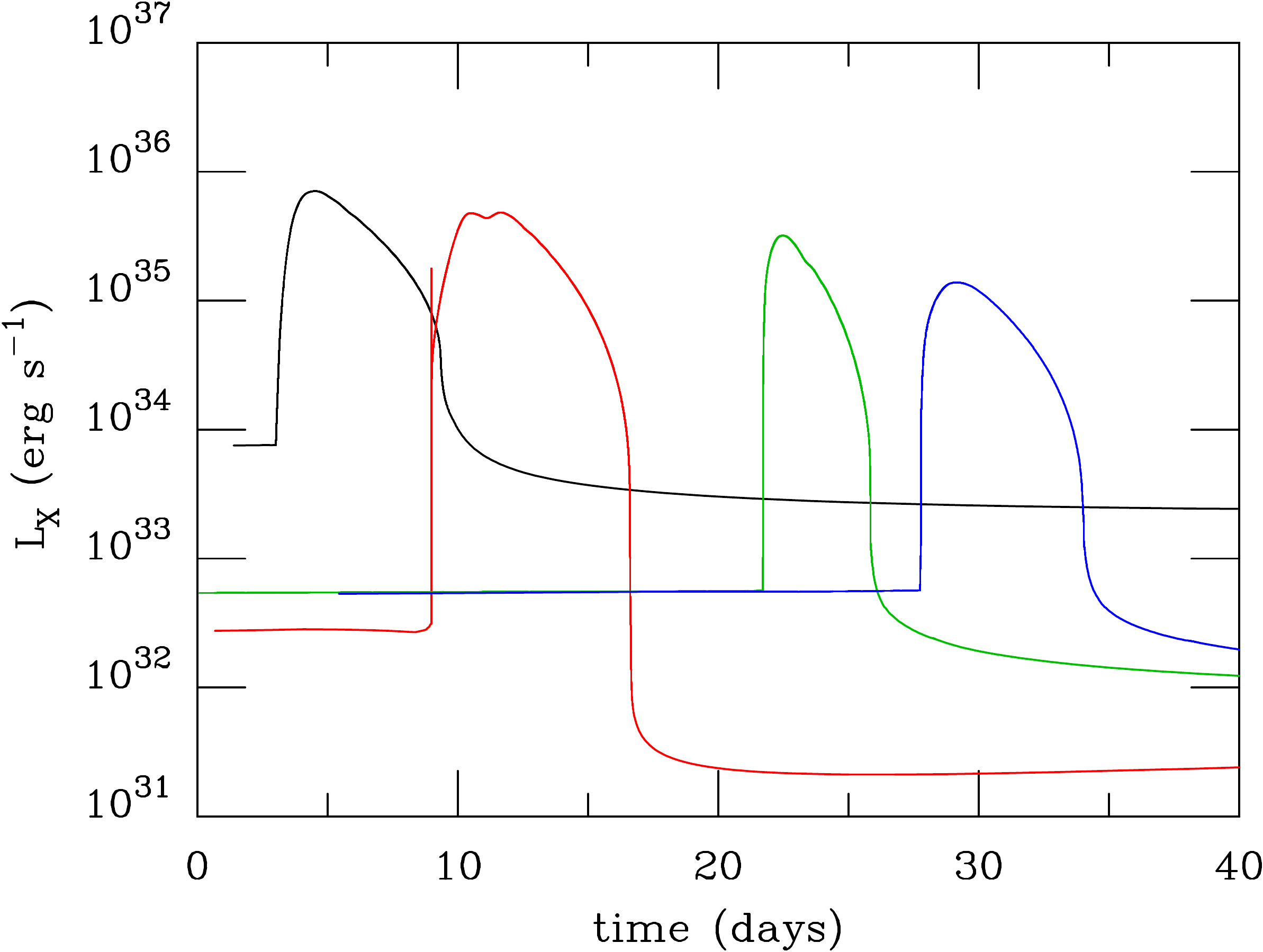}
   \caption{Zoom of the outbursts shown in Fig. \ref{lc_vfxb}. For a better legibility, the origin of the time axis is not the same as in Fig. \ref{lc_vfxb}.}
   \label{lc_vfxb_zoom}%
   \end{figure}

Faint outbursts are expected when either the accretion disc is very small or when the mass-transfer rate is so low that the entire disc cannot be brought into a hot state. Figure \ref{lc_vfxb} shows such a situation; the orbital parameters are those of Fig. \ref{lc-He}, but the mass-transfer rate is reduced to 10$^{14}$ gs$^{-1}$, with a variable or fixed small inner radius. A zoom of the outbursts is shown in Fig. \ref{lc_vfxb_zoom}.
   
The recurrence time between outbursts is 1.0 yr in the case of disc truncation, almost unchanged from the higher mass-transfer rate case (320 days), but the peak luminosity is reduced by more than one order of magnitude, and the duration is also shorter -- about 10 days. When the disc inner radius is set to $2 \times 10^8$ cm, the recurrence time is 180 days, slightly shorter than the value found for higher $\dot{M}$ (240 days). In both cases, a small fraction of the disc is accreted during an outburst (2.8\% in case of disc truncation, 2\% otherwise, to be compared to $\sim$30\% for the high $\dot{M}$ case), which is numerically challenging as one must follow more than 100 outbursts for the system to relax from the initial conditions. Interestingly, the outburst profile does not deviate much from an exponential; this is again due to the fact that the cooling front cannot propagate freely down to the inner disc radius, but instead a series of reflections between heating/cooling fronts occurs. The impact (in terms of rebrightenings) of these reflections on the light curve shape is, however, negligible from the observational point view.  The heating front never propagates beyond about 10$^{10}$ cm in both cases, and the outer disc remains in a cold, stable state. As for Fig. \ref{lc-He} and for the same reason, the quiescent level is much larger when the inner disc is truncated than when the disc inner radius is kept fixed at a small value. Note also that for the truncated disc, the mass accretion rate onto the neutron star is not much smaller than the mass-transfer rate from the secondary; the disc is therefore not very far from being stable on the cold branch, and 40\% of the mass transferred between outbursts is accreted onto the neutron star during quiescence. This helps in having a recurrence time significantly longer than in the case where the inner disc radius is set to $2 \times 10^8$ cm. This would support the idea that in these systems, as in bright transients \citep{dhl01,bz16}, accretion discs are truncated during quiescence, which could in principle be tested with high quality broad-band X-ray observations of these systems. While we do favour this hypothesis, one should keep in mind that other alternatives exist, because the viscosity parameter is not as constrained as in H-rich discs (see below).

Longer recurrence time can be obtained for lower values of $\alpha_{\rm c}$. Figure \ref{lc_vfxb} shows the case where $\alpha_{\rm c}$ is reduced to 0.01, for $\alpha_{\rm h}$ equal to 0.1 or 0.2. In both cases, the mass-transfer rate is $10^{13}$ g s$^{-1}$. The recurrence time is 2.8 yr for $\alpha_{\rm h}=0.2$ and 1.8 yr when $\alpha_{\rm h}=0.1$. The front propagates to a maximum distance of $4.8 \times 10^9$ cm in the first case, and to $4.1 \times 10^9$ cm in the second, and the outbursts are therefore extremely faint, with peak luminosities of $3.2 \times 10^{35}$ and $1.4 \times 10^{35}$ erg s$^{-1}$ respectively, and have short durations.

It is a general characteristic of the model that the faintest outbursts are also the shortest ones, simply because these correspond to the smallest extension of the hot, inner part of the disc. Longer outbursts could in principle be obtained by reducing the viscosity in the hot state $\alpha_{\rm h}$. Whereas there are good reasons to expect that $\alpha_{\rm c}$ should be different for discs in ultracompact binaries and in longer period systems, there is no obvious reason for such a change in $\alpha_{\rm h}$. The DIM would therefore have difficulties in explaining long and very faint outbursts. In particular those from \object{AX J1745.6-2901} lasting more than 80 weeks, or from \object{CXOGC J174538.0-290022} lasting more than 30 weeks \citep{dw10}. These long outbursts could still be explained by the DIM, but an additional ingredient, such as an increase of the mass-transfer rate due to the illumination of the secondary star is needed. This effect has been invoked to account for the long duration of superoutbursts in SU UMa stars \citep{hlw00,smak08}, even though its magnitude is unclear \citep[see e.g.][]{vh07,c15}.

   \begin{figure}
   \centering
   \includegraphics[angle=-90,width=\columnwidth]{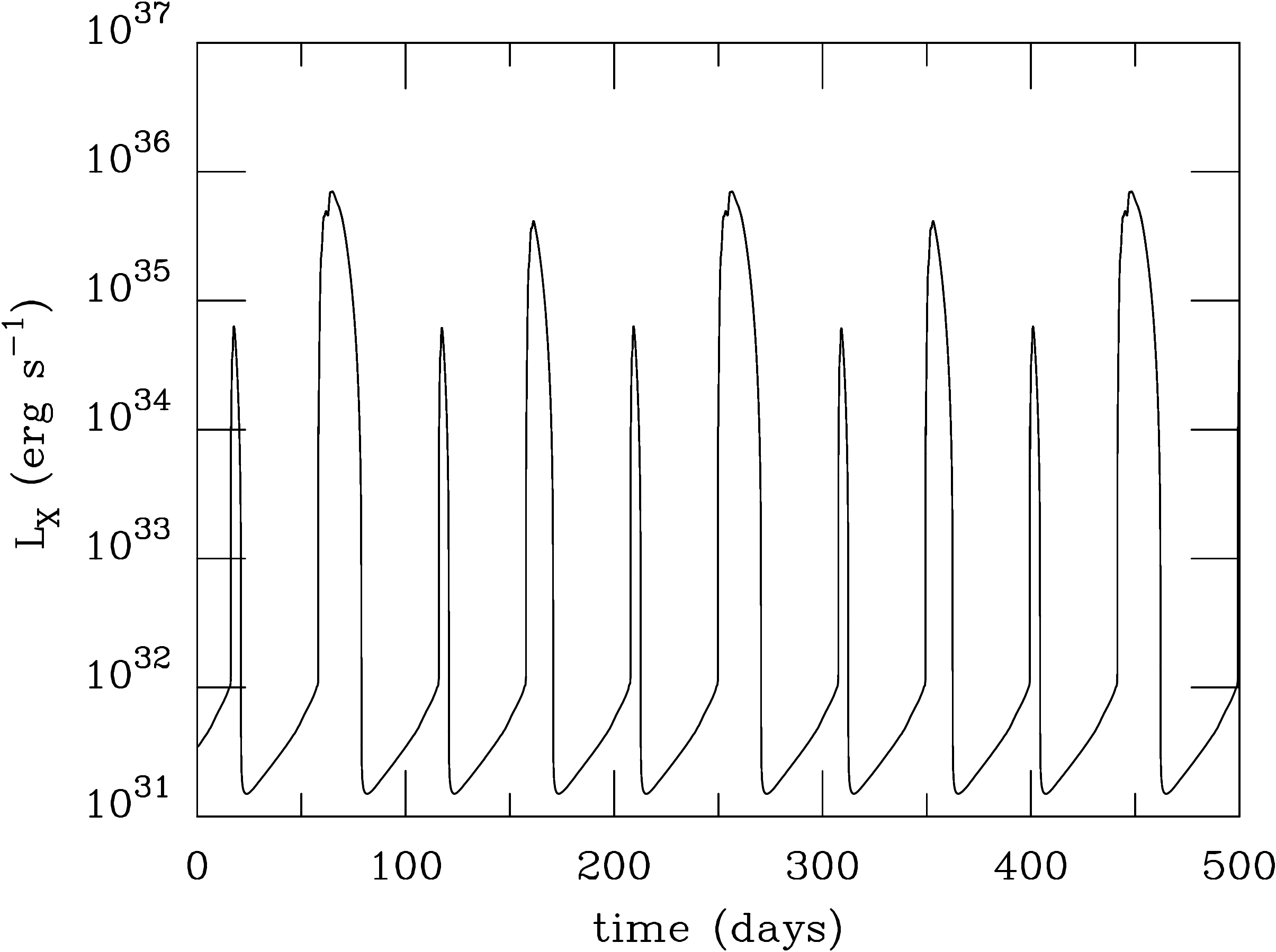}
   \caption{Outbursts profiles for low mass-transfer rates ( Here, $\dot{M}_{\rm tr} = 5 \times 10^{14}$ gs$^{-1}$), and for $\alpha_{\rm c} = 0.04$ and $\alpha_{\rm h}=0.2$. The disc is truncated with an inner disc radius given by Eq. (\ref{eq:rtrunc}).}
   \label{lc_vvfb}%
   \end{figure}
   
 We also considered the case where $\alpha_{\rm c}$ = 0.04, keeping $\alpha_{\rm h}$ unchanged to 0.2, which are standard values for cataclysmic variables and most probably for (helium) AM CVn stars \citep{KL12}. We have assumed here a tidal truncation radius of $1.0 \times 10^{10}$ cm, a mass-transfer rate of $5 \times 10^{14}$ g s$^{-1}$, and $C$ is kept equal to 0.01. The resulting profile is shown in Fig. \ref{lc_vvfb}; one can see sequence of very faint (10$^{35}$ erg s$^{-1}$) and faint (10$^{36}$ erg s$^{-1}$) outbursts.  As expected, the recurrence time is reduced as compared to the case with $\alpha_{\rm c} = 0.02$, and is of the order of 100 days between the main outbursts. The duration of faint outbursts is of the order of 10 days, that of the faintest ones is 4--5 days only; the entire disc is brought into the hot state during large outbursts, whereas the heating front can reach at most 40\% of the disc radius during the faintest ones. 
   
Interestingly, the peak luminosity of the faintest outbursts is close to that observed during the intermediate state of very faint X-ray transients, such as for example \object{XMM J174457-2850.3} \citep{dwr14}. During these states, the X-ray luminosity varies between typically $10^{33} - 10^{34}$ erg s$^{-1}$; they have been tentatively explained as resulting from the interaction of the accretion flow and the magnetic field of the neutron star. The extremely faint outbursts expected from the DIM would be very difficult to identify, and they would contribute to the variability observed during the intermediate state.

\section{The case of H discs}
 
  \begin{figure}
   \centering
   \includegraphics[angle=-90,width=\columnwidth]{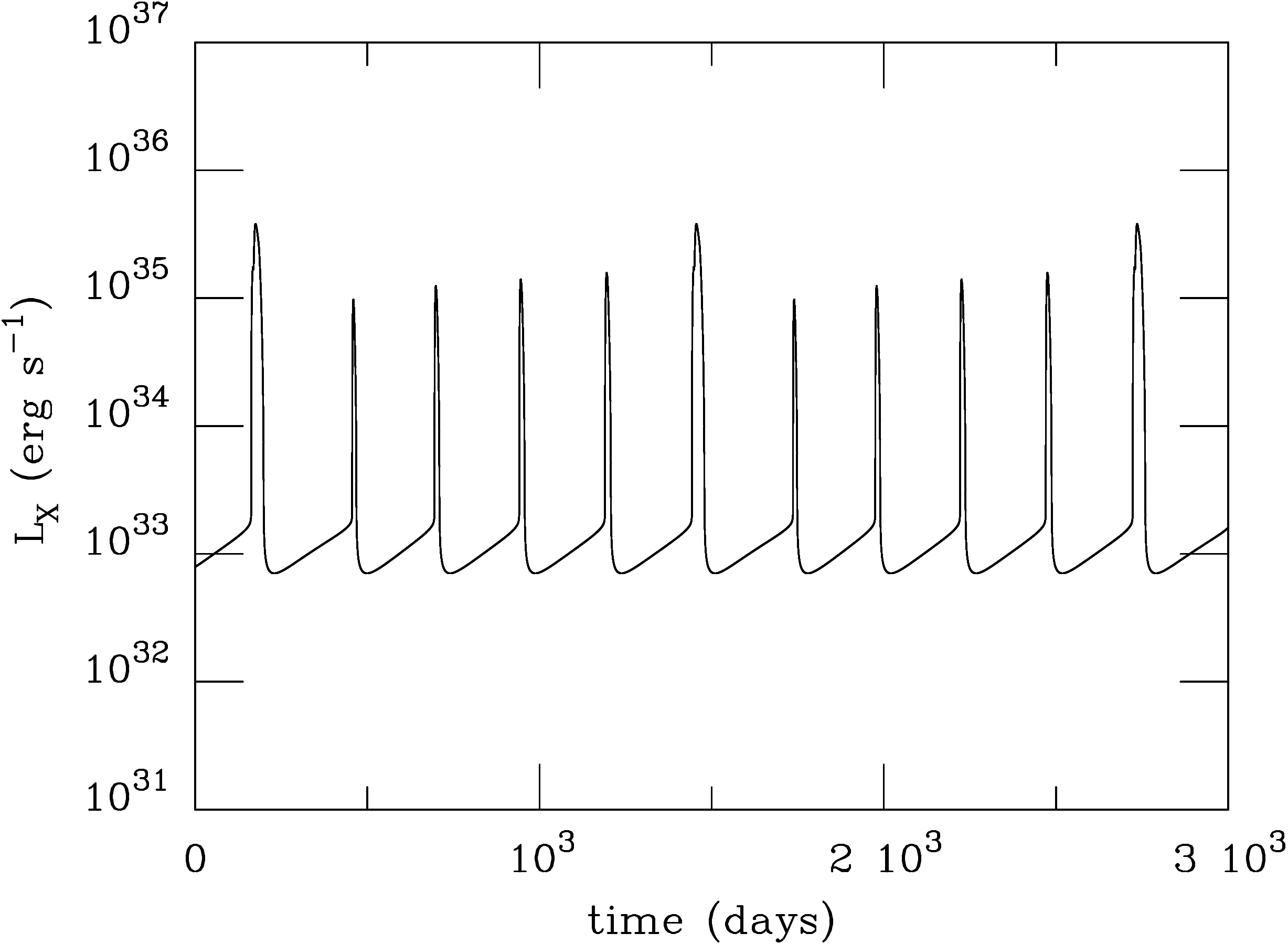}
   \caption{X-ray light curve in the case of a irradiated hydrogen disc. The orbital parameters and the viscosity are the same as in Fig. \ref{lc-He} and \ref{lc_vfxb}; the mass-transfer rate is $\dot{M}_{\rm tr} = 10^{14}$ gs$^{-1}$, again as in Fig. \ref{lc_vfxb} }
   \label{lc_h}%
   \end{figure}
   
Hydrogen rich companions are not expected in ultracompact binaries, as systems with H rich secondaries  should have a minimum orbital period around 80 minutes. It is, however, interesting to consider the case of H--rich secondaries, as the comparison with the He--rich case helps understanding the specificities of ultracompact binaries. They are also expected to behave in a similar way to systems containing a carbon-oxygen white-dwarf donor. Such systems should not be infrequent \citep[see e.g.][]{NJ10}, but modelling them is hampered by the unavailability of opacities for CO mixtures at all temperatures. \cite{mph01} have shown that accretion discs made of carbon and oxygen are subject to the same instability as hydrogen and helium discs, but have constructed thermal equilibrium curves only for pure carbon or pure oxygen above a few thousand degrees. Examining their Fig. 2 shows that the thermal instability for carbon and oxygen discs is triggered at about the same effective temperature as for hydrogen rich discs, which is not unexpected since the ionisation potential of those elements are similar, and much lower than that of helium.

Figure \ref{lc_h} shows the case of a hydrogen disc with the same parameters as for the black curve in Fig. \ref{lc_vfxb}, i.e. $M_1 = 1.4$ M$_\odot$, $M_2=0.035$ M$_\odot$, $P_{\rm orb}=0.7$ hr, and with a variable inner radius given by Eq. (\ref{eq:rtrunc}). We use $\alpha_{\rm c}= 0.02$  and $\alpha_{\rm h}= 0.2$, $C=0.01$, and $\dot{M}_{\rm tr} = 10^{14}$ gs$^{-1}$, also unchanged. Although it is impossible for a hydrogen rich secondary star to fit inside its Roche lobe for these orbital parameters, we chose not to change them to compare the effect of changing the disc composition only, and also because this particular model would be relevant for C/O degenerate secondaries if indeed carbon/oxygen discs do behave in a similar way as hydrogen rich discs. As can be seen the outbursts are significantly weaker than in the hydrogen poor case; they are so weak that the disc is always truncated, even at the outburst peak: the inner disc radius is always larger than 2.5 10$^{8}$ cm. The heating front reaches the outer disc edge only for the largest outbursts; for the weakest ones, the heating front does not propagate to distances larger than $1.1 \times 10^{10}$ cm. 

\section{Steady sources}

Several ultracompact X-ray binaries are steady \citep{hvn12}, and these are sources bright enough for their disc to sit entirely on the hot stable branch \citep{ldk08}. However, some VFXBs are also found to be steady with luminosities of order of a few times $10^{34} - 10^{35}$ erg s$^{-1}$\citep[see e.g.][]{icm05,dsm07,djt10,adw13}. These sources could in principle be  on the hot stable branch, provided that their orbital period is short enough. Assuming a primary mass of 1.4 M$_\odot$, and using Eqs. (12) and (14) from \citet{ldk08}, one finds that this happens for orbital periods shorter than 3.3 $L_{35}^{0.60}$ minutes in the case of helium discs and 13.2 $L_{35}^{0.63}$ min in the case of a mixed composition ($X=0.1$, $Y=0.9$), where $L_{35}$ is the X-ray luminosity in 10$^{35}$ erg s$^{-1}$ units. These orbital periods are extremely short and the mass-transfer rates expected from angular momentum losses due to gravitational radiation are orders of magnitude larger than the observed values.

The other option is that these sources stay on the cold stable branch. This is possible if the truncation radius is large enough, so that $\dot{M}_{\rm tr} < \dot{M}_{\rm crit}^-(r_{\rm in})$. For a pure helium disc, this happens when $r_{\rm in} > 3.1 \times 10^9 L_{35}^{0.38}$ cm. The situation would then be similar to that of \object{WZ Sge} for which \citet{hlh97} proposed that the source would stay on the cold stable branch during quiescence. WZ Sge outbursts would then be triggered if and when, for some reason, the mass-transfer rate increases above $\dot{M}_{\rm crit}^-(r_{\rm out})$. Mass transfer variations are observed in cataclysmic variables and LMXBs; for such a scenario to work one simply needs the average mass-transfer rate not to be much lower than $\dot{M}_{\rm crit}^-(r_{\rm out})$. \cite{lny96} and \cite{mnl99} suggested the existence of (almost) permanently quiescent X-ray binaries with large truncation radii. The first such system might have been recently found by \citet{tetarenkoetal16}.  If this scenario were also applicable to quiescent VFXBs, it would raise the possibility that these sources undergo very rare outbursts.

There are a few faint quasi-persistent sources which stay active for decades but which sometimes undergo deep quiescence with levels $< 10^{33}$ erg s$^{-1}$ \citep{dsm07,ash15}. In principle, these sources should not be on the hot branch during the quasi-steady phase, for reasons explained above. If they are on the cool branch during these quasi-steady states, the transition to very faint quiescence cannot be due to the disc instability. The only remaining possibility is a drop in mass transfer from the secondary. These drops, possibly due to large stellar spots, have been invoked in H-rich CVs to account for some peculiar systems \citep[see e.g.][]{hl14}, but it remains to be shown that this mechanism is applicable to very low-mass helium secondaries.

\citet{hbd15} suggested that the persistent VXFBs can explained by the propeller effect enabling only a small fraction of the accretion flow to reach the neutron star surface. The effects of the magnetic field is complex, and the ability of the propeller mechanism to produce the observed X-ray luminosity remains to be investigated (see below).
   
\section{Conclusions}
  
We have shown that the disc instability model applied to ultracompact binaries produces outbursts whose characteristics are very similar to those of the faint and very faint outbursts observed in VFXBs, provided we use the same value of the viscosity parameter $\alpha_{\rm h}$ as in cataclysmic variables and in longer period X-ray binaries, and a smaller (factor a few) value of $\alpha_{\rm c}$ in quiescence. This is not surprising because the non-ideal-MHD effects that are likely affecting the magnetically driven turbulence in quiescent, poorly ionised discs \citep[e.g.][]{Bai14,Lesuretal14}, might depend on their chemical composition \citep[see, however,][]{kld12}. In any case, contrary to $\alpha_{\rm h}$, whose value is well determined from observations which consistently produced a value of $0.1 - 0.2$ \citep[see e.g.][]{KL12}, the value of $\alpha_{\rm c}$ does not have the same status because its observational determination is less direct than that of $\alpha_{\rm h}$. Various estimates give values from 0.0001 to 0.04 \citep[see e.g.][]{lasota-01}. The outburst faintness results from the small size of the accretion disc and from the low mass-transfer rate. 

In the case of a 1.4 M$_\odot$ neutron star accreting from a helium degenerate secondary, the secular mean of the mass-transfer rate is \citep{hnv12}:
\begin{equation}
\dot{M}_{\rm tr}= 6.66 \times 10^{13} P_{\rm hr}^{-5.32} \; \rm g s^{-1} .
\end{equation}
For the 0.7 hr orbital period considered here, this gives $\dot{M}_{\rm tr} = 4.4 \times 10^{14}$ gs$^{-1}$. Of course, the actual mass-transfer rate from the secondary need not be equal to the secular mean, but it is worth noting that the values used in this paper are within factors $\sim$ a few equal to the secular mean. 

We also predict that many of these systems with low mass-transfer rates could exhibit two classes of outbursts: those where the heat front is able to bring the whole disc in a hot state and are therefore relatively bright and last longer than the much fainter ones for which the heating front is unable to reach the outer disc edge. It is possible that these extremely faint outbursts are not detected as such because of their faintness -- in order to increase the sensitivity of short {\it Swift} exposures, data had to be added within time frames of 2-4 weeks --, and that they could be partly responsible for the luminosity variations observed during the intermediate state of very faint X-ray transients such as \object{XMM J174457-2850.3} \citep{dwr14}. One should, however, notice that the duty cycle we produce is small, and that it would be difficult to account for all of the observed luminosity variations observed during the intermediate state. 

Very faint outbursts can also be produced in long period binaries, provided that the mass-transfer rate is low enough. For the $\dot{M}_{\rm tr} = 10^{13}$ gs$^{-1}$ cases presented in Section 2.2.2, the heating front never reached distances larger than a third of the disc size, and the outer parts of the disc were therefore in a steady state on the cool branch; the outburst properties are therefore independent of the disc size, and the same light curves would have been obtained for the same mass-transfer rate for longer orbital periods.

Very faint outbursts have been observed in several bright transients, such as for example \object{Aql X-1} \citep{ccd14}, \object{XTE J1701-462} \citep{f10}, \object{KS 1741-293} \citep{dw13} or SAX J1750.8-2900 \citep{wd13}. The possibility that these transients are of the same nature as those mentioned in this paper, with heat-front not able to propagate throughout the disc is interesting, but needs to be investigated in more details. This is clearly out of the scope of this paper.

One should also notice that the truncated helium disc that we have considered here is stable when the mass-transfer rate is lower than $\dot{M}_{\rm crit}^-(r_{\rm in})$; for the parameters we have used here, this corresponds to X-ray luminosities of $10^{33} - 10^{34}$ erg s$^{-1}$. Such stable low luminosity systems have been observed in the case of AM CVn systems \citep[see e.g.][and references therein]{Bildstenetal06} and are predicted by the DIM \citep{smak83,kld12}.

For a given orbital period, one therefore obtains the following sequence for decreasing the mass-transfer rates: for high $\dot{M}_{\rm tr}$, the system is stable on the hot branch; at lower $\dot{M}_{\rm tr} < \dot{M}_{\rm crit}^+(r_{\rm out})$, (relatively) bright outbursts are observed, during the entire disc is being brought in the hot state; for lower $\dot{M}_{\rm tr}$, one obtains a sequence of bright and faint outbursts during which the heating front does not reach the outer disc edge, and for still lower $\dot{M}_{\rm tr}$, only faint outbursts are obtained. Finally, for $\dot{M}_{\rm tr} < \dot{M}_{\rm crit}^-(r_{\rm in})$, the system is stable on the cold branch. In all cases, the outbursts are expected to be short, and depend mainly on $\alpha_{\rm c}$, and, for the outbursts able to reach the outer disc edge, on the orbital period. In all cases, faint outbursts are expected to be short if no additional ingredient, such as an increase of the mass-transfer rate due to the illumination of the secondary, is added.

We have not taken into account the interaction of the neutron star magnetic field with the accreted matter. This magnetic field  could be the reason for the disc truncation; it could also prevent accretion onto the neutron star via the propeller effect, thereby accounting for the observed transition between the millisecond pulsar state and the LMXB state \citep[see e.g.][]{pt15}. However, the recent observation of X-ray pulsations in the accreting millisecond X-ray pulsar \object{PSR J1023+0038} while the system was in quiescence with an X-ray luminosity of $3 \times 10^{33}$ erg s$^{-1}$ \citep{abp15} or of \object{XSS J12270-4859} which has similar properties \citep{pmb15} shows that even at those low mass--accretion rates, the propeller mechanism does not always prevent accretion. It is also worth noting that the light cylinder radius, equal to $4.8 \times 10^6 P_{\rm ms}$ cm, where $P_{\rm ms}$ is the spin period of the millisecond pulsar, is smaller than the truncation radius given by Eq. (\ref{eq:rtrunc}) for millisecond rotation periods. The effect of the neutron star magnetic field is thus more complex than usually assumed, and it is unlikely that simple recipes can be used to decide if the magnetic field is able to disrupt the accretion disc or not.

Finally, one should keep in mind that the neither the VFXB nor the very faint X-ray transient categories are homogeneous. Apparently not all VFXBs are ultracompact and probably not all very faint outbursts can be explained by the DIM in ultracompact binaries.

\begin{acknowledgements}
We thank the anonymous referee for very useful comments which helped to improve this paper. This work was supported by a National Science Centre, Poland grant 2015/19/B/ST9/01099. JPL was supported by a grant from the French Space Agency CNES.
\end{acknowledgements}

\end{document}